# Giant orbit-to-charge conversion induced via the inverse orbital Hall effect

Renyou Xu,[1,2] Hui Zhang,[1*] Yuhao Jiang,[1] Houyi Cheng,[1,2] Yunfei Xie,[3] Yuxuan Yao,[1] Danrong Xiong,[1] Zhaozhao Zhu,[4,5] Xiaobai Ning,[1] Runze Chen,[1] Yan Huang,[1] Shijie Xu,[1,2] Jianwang Cai,[4] Yong Xu,[1,2] Tao Liu,[3] Weisheng Zhao[1,2*]

[1]Fert Beijing Institute, School of Integrated Circuit Science and Engineering, Beihang University, Beijing 100191, China

[2]Hefei Innovation Research Institute, Beihang University, Hefei 230013, China

[3]National Engineering Research Center of Electromagnetic Radiation Control Materials, University of Electronic Science and Technology of China, Chengdu 610054, China

[4]Beijing National Laboratory for Condensed Matter Physics, Institute of Physics, Chinese Academy of Sciences, Beijing 100190, China

[5]Songshan Lake Materials Laboratory, Dongguan, Guangdong 523808, China

*Corresponding authors:
weisheng.zhao@buaa.edu.cn;
huizh@buaa.edu.cn;
These authors contributed equally: Renyou Xu, Hui Zhang, Yuhao Jiang, Houyi Cheng

Abstract:

We investigate the orbit-to-charge conversion in YIG/Pt/nonmagnetic material (NM) trilayer heterostructures. With the additional Ru layer on the top of YIG/Pt stacks, the charge current signal increases nearly an order of magnitude in both longitudinal spin Seebeck effect (SSE) and spin pumping (SP) measurements. Through thickness dependence studies of the Ru metal layer and theoretical model, we quantitatively clarify different contributions of the increased SSE signal that mainly comes from the inverse orbital Hall effect (IOHE) of Ru, and partially comes from the orbital sink effect in the Ru layer. A similar enhancement of SSE(SP) signals is also observed when Ru is replaced by other materials (Ta, W, and Cu), implying the universality of the IOHE in transition metals. Our findings not only suggest a more efficient generation of the charge current via the orbital angular moment channel but also provides crucial insights into the interplay among charge, spin, and orbit.

The observation of the orbital torque generated by the orbital Hall effect (OHE) and orbital Rashba-Edelstein effect (OREE), has recently drawn a great deal of attention due to the utilization of orbital angular moment (OAM) as a new information carrier for future information technology [1-3]. In the scenario of OHE, a flow of OAM or nonequilibrium OAM accumulation, whose orbital polarization is transverse to the charge current, can be generated in transition metals independent of strong spin-orbit coupling (SOC), which is a bulk effect [4,5]. The confirmation of charge-to-orbit conversion, as evidenced by the orbital torque switching of magnetization and theoretical calculations, provides a strong indication of the possibility of orbit-to-charge conversion, considering the Onsager reciprocity [6,7]. It has been suggested that OHE is significantly stronger than the spin Hall effect (SHE) by an order of magnitude in most transition metal materials [1]. Consequently, we anticipate that the inverse orbital Hall effect (IOHE) is considerably stronger than the inverse spin Hall effect (ISHE), which could fundamentally contribute to the development of memory storage devices and enhance the efficiency of ISHE-based technologies, such as magnetoelectric spin-orbit (MESO) logic [8,9]. Over the past years, plenty of works have been devoted to exploring the ISHE [10-13]. However, thus far, only a few studies have explored the IOHE through the spin terahertz measurements [14,15]. The film thickness-dependent absorption of the terahertz signal, coupled with spin/orbit accumulation induced by ultrafast demagnetization, makes it difficult to disentangle the contributions of ISHE and IOHE [16-19]. To unlock the underlying physics, it is essential to employ alternative spin/OAM injection methods and perform theoretical analyses encompassing the interactions between charge, spin, and orbital degrees of freedom.

In this work, we demonstrate the unambiguous detection of IOHE in YIG/Pt/NM heterostructures through longitudinal spin Seebeck effect (SSE) and spin pumping (SP), where NM represents nonmagnetic material. Remarkably, a substantial enhancement in SSE and SP signals was observed upon depositing a light metal, Ru, on top of the YIG/Pt sample. Furthermore, the signal reached saturation with increasing thickness of the Ru layer, as confirmed by the measurements of the thickness dependence of orbit-to-charge conversion. Theoretical analysis based on a spin-orbit diffusion model exhibited good agreement with the experimental results, indicating that the IOHE of Ru primarily contributes to the increased SSE signal. Notably, our findings demonstrate the existence of an orbital sink effect in the Ru layer, which partially promotes the SSE signal. A similar enhancement of SSE(SP) signals is also observed when Ru is replaced by Ta, W, and Cu, suggesting that the IOHE is a universal phenomenon in these materials. These results shed light on the mystery of the IOHE and offer a promising avenue for the development of spin-orbitronics devices based on the IOHE.

YIG/Pt/NM trilayer heterostructures were prepared using DC/RF magnetron sputtering with the basic pressure of the chamber below $10^{-8}$ mbar [20]. A 40-nm-thick epitaxial $Y_3Fe_5O_{12}$ (YIG) film was deposited on a (111)-oriented gadolinium gallium garnet (GGG) single crystal substrate. A slight roughness of 0.078 nanometers of the YIG layer was obtained through Atomic Force Microscopy (AFM) (see S1 in the Supplemental Material for details [21]). We further confirmed the high structural quality of the YIG layer by X-ray diffraction (XRD). Fig. 1(a) shows the representative $\theta$-$2\theta$ scan of the GGG/YIG heterostructure, demonstrating the purity of the YIG phase. The appearance of Laue oscillations near the YIG (444) diffraction peak provides evidence of the film's exceptional uniformity. Fig. 1(b) displays the hysteresis loop of the GGG/YIG sample under the in-plane magnetic field obtained by a Lake Shore vibrating sample magnetometer (VSM) at room temperature. The GGG/YIG sample demonstrates a saturation magnetization of 95.1 emu/cc, with a coercive field below 2 Oe, which agrees well with the previous work [22,23]. The thin film multilayers consist of Pt(1.5)/NM($t_{NM}$)/MgO(3)/Ta(2) ($t_{NM}$ denotes the thickness of the NM layer, and numbers indicate the thickness in nm) deposited on top of the GGG/YIG sample (see S1 in the Supplemental Material for details [21]). MgO(3)/Ta(2) serves as the capping layer to prevent further oxidation of the NM layer, which will be omitted in electric transport experiments for convenience. Since the thin Ta layer exhibits high resistivity when exposed to the atmosphere, the current-shunting effect induced by the capping layer was negligible in our experiments.

Fig. 1(c) presents the cross-section of the YIG(40)/Pt(1.5)/Ru(6)/MgO(3)/Ta(2) stack, as captured by high-resolution transmission electron microscopy (HR-TEM). The YIG layer, MgO layer, and Ru layer exhibit well-crystallized structures. Fig. 1(d) showcases the corresponding energy dispersive spectroscopy (EDS) images, revealing the distribution of Y, Pt, Ru, and Mg elements, respectively. The interdiffusion of elements between different layers is minimal, as no additional annealing process was applied. The HR-TEM and EDS results validate the high quality of the sample and the sharp interfaces between the different layers.

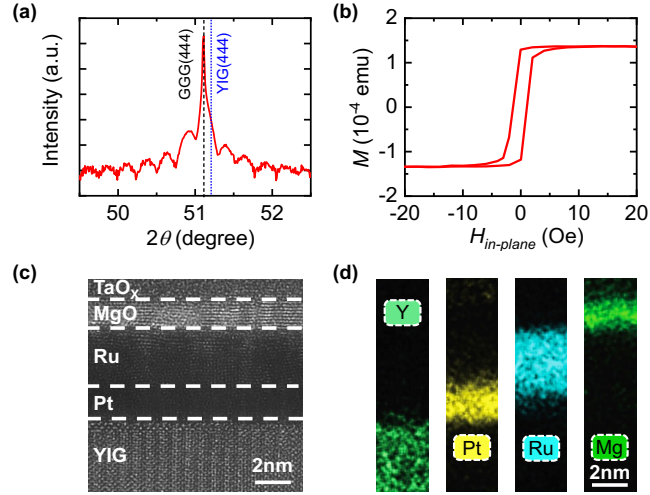

FIG. 1. Structure and magnetism characterization. (a) The XRD pattern of the GGG/YIG sample. (b) The magnetic loop of the GGG/YIG sample under the in-plane magnetic field. (c) The cross-sectional HR-TEM image of the YIG(40)/Pt(1.5)/Ru(6)/MgO(3)/Ta(2) multilayer. (d) The EDS mapping of corresponding elements.

We carry out longitudinal SSE measurements in a Quantum-designed physical property measurement system (PPMS), and a sketch for the experiment setup is shown in Fig. 2(a). The sample was thermally connected to a heater, with its temperature measured via a thermocouple. At the same time, a copper block served as a heat sink at the top of the sample surface, which was thermally contacted to the chamber of PPMS, maintaining a constant temperature of 300 K. A temperature difference, denoted as $\Delta T$, was established along the z-axis of the film, with most of the $\Delta T$ occurring in the thicker YIG layer due to its lower thermal conductivity compared to the metal layers. The temperature difference $\Delta T$ within the YIG layer induces a pure spin current $S_{heat}^{YIG}$, which is then injected into the Pt layer. The spin index σ aligns parallel to the magnetization of the YIG layer, and its orientation can be controlled by applying an in-plane magnetic field. As a result of the ISHE, the charge current $I_{Pt}$ generated in the Pt layer induces an electric field in the direction of σ×$S_{heat}^{YIG}$, giving rise to a voltage denoted as $V_{SSE}$ along the x-axis [24].

Fig. 2(b) displays the SSE signals $I_{SSE}$ of YIG(40)/Pt(1.5) (black) and YIG(40)/Pt(1.5)/Ru(4) (red) as a function of the magnetic field $H_y$ along the y-axis. The charge current $I_{SSE}$ was defined as $V_{SSE}/R$, where the voltage $V_{SSE}$ was measured by a Keithley 2182 nanovoltmeter, and $R$ corresponds to the resistance of the electrical contacts between the sample and the nanovoltmeter. The magnetic field swept between -20 Oe and 20 Oe, and the heating power was fixed at 300 mW to maintain a constant temperature difference (ΔT =13 K) for both samples during the test. By adding a 4-nm-thick Ru layer on the top of the YIG(40)/Pt(1.5) bilayer, the $I_{SSE}$ dramatically increased nearly an order of magnitude[Fig. 2(b)], rising from 0.17 nA to 1.58 nA. We modulated the temperature difference by adjusting the heating power. With the increased heating power, a larger temperature difference leads to a larger injected spin current, and a significant enhancement of the $I_{SSE}$ was observed (see S2 in the Supplemental Material [21]). Considering the weak SOC of Ru, ISHE in the Ru layer is much smaller compared to ISHE in the Pt layer [25-27]. As a result, the traditional ISHE theory cannot explain the significant SSE signal enhancement.

Considering the negligible ISHE of Ru, we argued that the plausible mechanism to explain this phenomenon is the occurrence of IOHE in the Ru layer. Despite the quenching of OAM in transition metal, OHE leads to the local accumulation of OAM due to the applied transverse electric field, which has been proven by the Kerr effect and orbital torque effect experiments recently [28-30]. Analogous to the ISHE that can convert spin current into charge current, the IOHE enables the conversion between OAM and the charge current, which means the injection of an orbital current will leads to induced charge current in a direction that is perpendicular to the orbital current and orbital polarization. The orbital current can be understood as a wave packet carrying OAM, which can be induced through OHE and OREE [31,32].

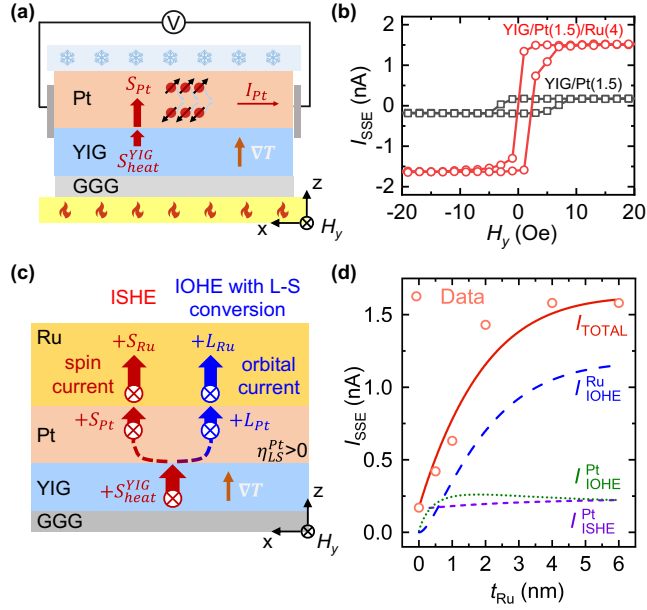

FIG. 2. SSE measurements. (a) Longitudinal SSE in YIG/Pt bilayer. (b) A comparison of the SSE signals between the YIG(40)/Pt(1.5)/Ru(6) (red) and YIG(40)/Pt(1.5) (black) is presented. These signals were obtained using a heating power of 300 mW. (c) Longitudinal SSE in YIG/Pt/Ru trilayer heterostructures considering spin-to-orbit conversion (L-S conversion), ISHE, and IOHE. (d) SSE signal of YIG(40)/Pt(1.5)/Ru($t_{Ru}$) samples as a function of the thickness of the Ru layer (red circle). Here the purple dotted line, green dashed line, and blue dashed line indicate fitting curves of the SSE signal with contributions from charge current induced by the ISHE of Pt ($I_{ISHE}^{Pt}$), the IOHE of Pt ($I_{IOHE}^{Pt}$), and the IOHE of Ru ($I_{IOHE}^{Ru}$), respectively. The ISHE of Ru ($I_{ISHE}^{Ru}$) is much weaker compared to the other contributions, thus not shown in this figure. The solid red line represents the fitting curves $I_{TOTAL}$ with all four contributions mentioned above.

In consideration of the large orbital Hall conductivity of Ru ($\sigma_{OH}^{Ru}$ = 9100 (ℏ/e) (Ω cm)$^{-1}$) that was observed in recent works [33-35], we propose a physical mechanism as shown in Fig. 2 (c). The temperature difference drives spin current $S^{Pt}$ diffuses into the Pt layer, a transverse charge current $J_C$ ($J_C = \frac{2e}{\hbar}\frac{\sigma_{SH}^{Pt}}{\sigma^{Pt}}S^{Pt}$, where $\sigma_{SH}^{Pt}$ and $\sigma^{Pt}$ is the spin Hall conductivity and conductivity of Pt) is generated as a result of ISHE. At the same time, an orbital current $L^{Pt}$ ($L^{Pt} = \eta_{LS}^{Pt}S^{Pt}$, where $\eta_{LS}^{Pt}$ is the conversion coefficient of spin-to-orbit for Pt) generates in the Pt layer due to the strong SOC of Pt and then diffuses into the Ru layer. Based on the scenario of IOHE, the orbital current $L^{Ru}$ can also be converted to a transverse charge current $J_C$ ($J_C = \frac{2e}{\hbar}\frac{\sigma_{OH}^{Ru}}{\sigma^{Ru}}L^{Ru}$, where $\sigma^{Ru}$ is the conductivity of Ru) in the Ru layer. Since $\sigma_{OH}^{Ru}$, $\eta_{LS}^{Pt}$, and $\sigma_{SH}^{Pt}$ have the same positive sign, the additional Ru layer shall enlarge the SSE signal of the YIG/Pt bilayer. To further understand the properties of the observed SSE signal enhancement, we conducted the Ru layer thickness $t_{Ru}$-dependent SSE measurements. As shown in Fig. 2(d), the SSE signal of YIG(40)/Pt(1.5)/Ru($t_{Ru}$) samples increases with $t_{Ru}$ increases and then reached saturation when $t_{Ru}$ = 4 nm (see S3 in the Supplemental Material for details [21]).

To better understand the contributions to the SSE signal, we have developed a one-dimension spin-orbit diffusion model (refer to Supplemental Material S4 for detailed calculations [21]), which contains the spin-to-orbit conversion in Pt due to its strong SOC [2,36,37]. As illustrated in Fig. 2(d), the fitting curves $I_{TOTAL}$ (red solid line) exhibit a good agreement with our experimental results (red circle), where $I_{TOTAL}$ encompasses four distinct SSE contributions involving ISHE and IOHE in both Pt and Ru layers, expressed as $I_{TOTAL} = I_{IOHE}^{Ru} + I_{ISHE}^{Ru} + I_{ISHE}^{Pt} + I_{IOHE}^{Pt}$. As the thickness of the Ru layer increases, the $I_{IOHE}^{Ru}$ (blue dashed line) experiences a rapid increase, ultimately becoming the predominant factor among all the contributions, which indicates the increased SSE signal mainly comes from the IOHE of Ru. The $I_{ISHE}^{Ru}$ is nearly constant at zero due to its weak SOC, thus is negligible in our experiment. As the spin current permeates through the Pt layer, a portion of this spin current reflects at the edge of the Pt layer, compensating $I_{ISHE}^{Pt}$. Consequently, introducing an extra Ru layer restrains the spin backflow, resulting in an increase of $I_{ISHE}^{Pt}$[38,39]. This phenomenon is evident in the slight elevation of the $I_{ISHE}^{Pt}$ (purple dotted line) as the Ru layer's thickness increases, denoting a spin sink influence (see S5 in the Supplemental Material for detail). However, the rise in $I_{ISHE}^{Pt}$ is notably minuscule in comparison to $I_{IOHE}^{Ru}$, suggesting that the influence of the spin sink effect can also be disregarded in our experiment. Furthered, $I_{IOHE}^{Pt}$ (green dashed line) exhibit an obvious increase and even exceed the $I_{ISHE}^{Pt}$ at the thicker Ru layer. We note that this phenomenon can be explained by an orbital sink effect. Since the considerable orbital current is generated and diffusion in the Pt layer, partial orbital current shall be reflected at the edge of the Pt layer and compensate for the IOHE that occurred in the Pt layer (see S5 in the Supplemental Material for detail). However, the additional Ru layer restrains the orbital backflow and boosts the orbit-to-charge conversion occurring in the Pt, thereby promoting the SSE signal[5,21,40,41]. By discerning between different contributions, we demonstrate that the impact of the spin sink of Ru appears to be negligible in our model, but the IOHE of Ru is primarily responsible for the increased SSE signal, while the orbital sink effect partially enhances the SSE signal.

We also performed spin injection measurements using SP driven by ferromagnetic resonance (FMR) to validate our findings from the longitudinal SSE measurements. Fig. 3(a) illustrates the experimental setup for the SP measurements. Upon RF excitation, a spin current is initially injected into the Pt layer, which then converts into an orbital current that diffuses into the Ru layer. A similar process is expected to occur as in the longitudinal SSE experiment, where both ISHE and IOHE contribute to the charge current signal in the YIG/Pt/Ru heterostructure. The detailed experiment setup was similar to our previous work [42]. Fig. 3(b) shows the measured SP signal of YIG(40)/Pt(1.5) sample as a function of the magnetic field under different frequencies, exhibiting a symmetrical Lorentzian shape. For a clear comparison, the charge current $I_{SP}$ is defined as $U_{SP}/R$, where $U_{SP}$ is directly measured by the lock-in amplifier, and $R$ is the electric resistance of the sample contacts. Upon reversing the external magnetic field (The angle between the external DC magnetic field and the x-axis ($\theta_H$) changes from 90° to 270°), the measured signal shows an opposite sign with an almost unchanged magnitude, indicating that the spin rectification effects are negligible in our experiment [43]. Fig. 3(c) displays the SP signals of YIG(40)/Pt(1.5) (red) and YIG(40)/Pt(1.5)/Ru(4) (black), both measured at 8 GHz. It can be observed that the $I_{SP}$ of YIG(40)/Pt(1.5)/Ru(4) is over 10 times higher than that of YIG(40)/Pt(1.5). Further information about the IOHE was obtained through a Ru thickness dependence experiment, as shown in Fig. 3(d). The $I_{SP}$ first increases as the Ru layer's thickness increases, then saturates at about 4 nm (see S5 in the Supplemental Material for details [21]). These results support the conclusion obtained by SSE measurements that the additional light metal Ru layer strongly enlarges the charge signal, which can be well explained by IOHE.

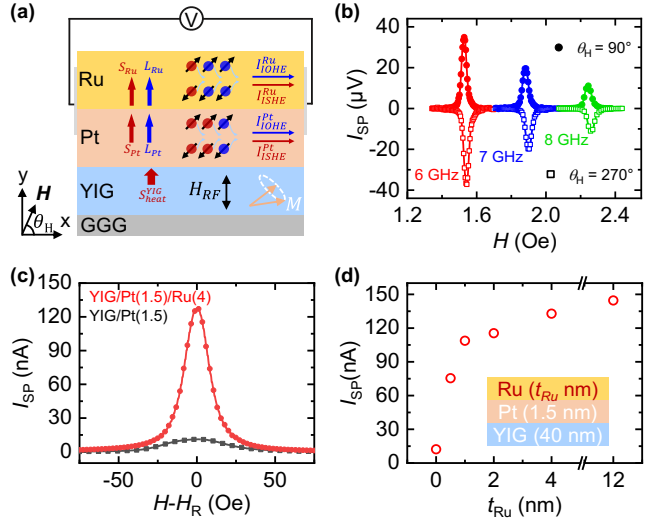

FIG. 3. SP measurements. (a) An illustration of FMR-driven SP. $H$ is the in-plane external DC magnetic field, and $H_{RF}$ is the out-of-plane radiofrequency magnetic field. Voltage is measured along the x direction. (b) Representative SP signal of YIG(40)/Pt(1.5). $\theta_H$ is 90° or 270° which denotes the angle between the external DC magnetic field and the x-axis. (c) SP charge current of YIG(40)/Pt(1.5) (black) and YIG(40)/Pt(1.5)/Ru(4) (red) samples measured at 8 GHz. (d) SP charge current as a function of Ru thickness in YIG(40)/Pt(1.5)/Ru($t_{Ru}$) samples measured at 8 GHz.

The Ru element exhibits a large $\sigma_{OH}$ compared to other transition metals, as evidenced by recent studies [5,44]. To corroborate the hypothesis that the dominant factor responsible for signal enhancement is the IOHE, additional experiments were conducted involving other transition metals possessing large $\sigma_{OH}$ [1,15]. We investigate the IOHE in the YIG/Pt(1.5)/NM(2) heterostructures (NM = Ta, W, Cu, and Ru) via both SSE and SP experiments, as shown in Figs. 4(a) and 4(b). The results are ordered from weak to strong signals, and it is evident that the sample with the Ru layer exhibits the largest signal. In addition, Figs. 4(a) and 4(b) also demonstrate the consistent agreement between SSE and SP measurements. Surprisingly, the signals of samples with NM=Ta and W also show a significant increase, which contradicts the theory of ISHE. According to ISHE, the SSE and SP signals generated in Ta and W should cancel out the Pt signal due to the opposite signs of the spin Hall conductivity ($\sigma_{SH}$) compared to Pt. However, the $\sigma_{OH}$ of Ta, W, and Pt have the same positive sign. Consequently, the IOHE overcomes the cancellation effect of ISHE in Ta and W, resulting in an increased signal. Furthermore, the sample of Cu, which is a light metal and a poor spin sink material (spin diffusion length over 200 nm measured at room temperature [45]), exhibits a substantial increase, supporting the notion that the spin sink effect is not prominent in our experiments.

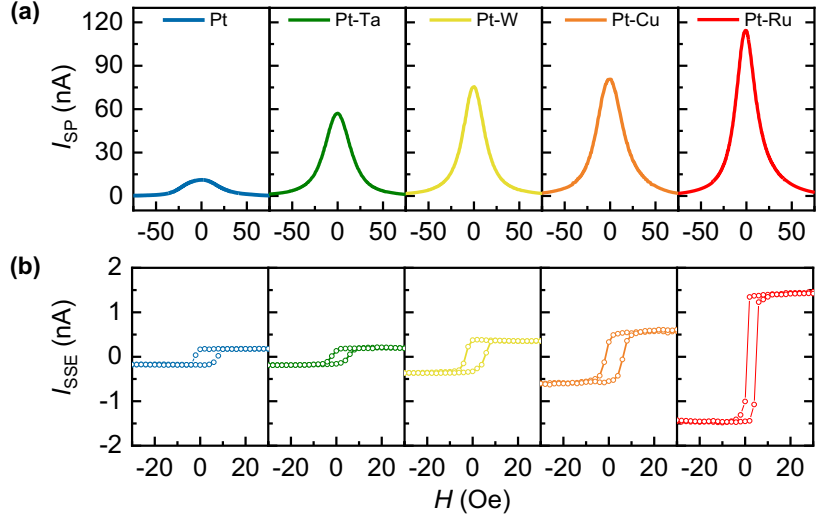

FIG. 4. (a) SP measurements and (b) SSE measurements of YIG/Pt/NM samples with different NM (NM = Ta, W, Cu, and Ru).

TABLE I. The sign of the $I_{ISHE}$ and $I_{IOHE}$ from the NM layer in YIG/Pt(Gd)/NM heterostructures.

|  |  | NM$_1$ | NM$_2$ | NM$_3$ |
|---|---|---|---|---|
|  | $\sigma_{SH}^{NM}/\sigma_{OH}^{NM}$ | +/+ | -/+ | +/- |
| $\eta_{LS}^{Pt}>0$ | $I_{ISHE}/I_{IOHE}$ | +/+ | -/+ | +/- |
| $\eta_{LS}^{Gd}<0$ | $I_{ISHE}/I_{IOHE}$ | +/- | -/- | +/+ |

In YIG/Pt/NM trilayer heterostructures, Pt serves as an efficient spin-to-orbit converter which enables the orbital current to inject into the NM layer, and promotes the orbit-to-charge conversion in the NM layer. We note that the SSE(SP) signal of a single NM layer in YIG/Pt/NM has a strong correlation with its $\sigma_{SH}$, $\sigma_{OH}$, and $\eta_{LS}^{Pt(Gd)}$. While recent work introduces the rare earth element Gd have stronger SOC enables the conversion between spin and orbital current. We compare ISHE and IOHE of different NM layers in YIG/Pt(Gd)/NM heterostructures SSE(SP) measurements, which is summarized in TABLE I. By assuming a positive spin current initially diffuses into the Pt layer, the utilization of Pt as a positive spin-to-orbit converter ($\eta_{LS}^{Pt}>0$) facilitates the injection of positive spin and orbital currents into the NM layer. Consequently, NM$_1$ with positive $\sigma_{OH}$ and $\sigma_{SH}$, such as Ru, exhibit stronger signals, while NM$_2$(NM$_3$) with opposite signs of $\sigma_{SH}$ and $\sigma_{OH}$ offset the signals. On the other hand, employing Gd as a negative spin-to-orbit converter ($\eta_{LS}^{Gd}<0$) leads to the injection of positive spin and negative orbital currents [4]. In such cases, the stronger signals are expected to obtain by using NM$_2$(NM$_3$) with large $\sigma_{SH}$ and $\sigma_{OH}$ but oppositive signs, such as transition metals (NM$_2$ = Ta and W), two-dimensional electron gas materials (NM$_2$ = 2DEG at the KTaO$_3$ or SrTiO$_3$ interface [46]), as well as transition metal disulfides (NM$_3$ = MoTe$_2$ and WTe$_2$ [47,48]).

In conclusion, this study presents a novel efficient approach to generating charge current utilizing the OAM channel. It is found that the Ru layer with negligible SOC can significantly boost the inverse charge signal through the orbit-to-charge conversion, as evidenced by the SSE and SP measurements. In addition to the previous studies of IOHE, our work combines thickness-dependent measurements with theoretical analysis, disentangles IOHE from ISHE, and surprisingly observes an orbital sink effect. These results shed light on the interaction between charge, spin, and orbit, offering potential benefits for advancing emerging spin-orbitronics applications such as MESO and spin terahertz emitters.

The authors thank Albert Fert for his useful discussion. The authors gratefully acknowledge the National Key Research and Development Program of China (No. 2022YFB4400200), National Natural Science Foundation of China (No. 92164206, 52261145694, 52072060, and 52121001), Beihang Hefei Innovation Research Institute Project (BHKX-19-01, BHKX-19-02). All the authors sincerely thank Hefei Truth Equipment Co., Ltd for the help on film deposition. This work was supported by the Tencent Foundation through the XPLORER PRIZE.